\documentclass[nofootinbib,letterpaper,floatfix,superscriptaddress,preprintnumbers,twocolumn]{revtex4}

\usepackage{color}

\usepackage{graphicx}

\usepackage{lipsum}
\makeatletter
\def\ps@pprintTitle{%
 \let\@oddhead\@empty
 \let\@evenhead\@empty
 \def\@oddfoot{}%
 \let\@evenfoot\@oddfoot}
\makeatother

\usepackage{verbatim}
\usepackage{amsmath}
\usepackage{amssymb}
\usepackage{subfigure}
\usepackage{url}

\usepackage{multirow}
\usepackage{threeparttable}
\usepackage{paralist}

\usepackage{color}
\definecolor{darkgreen}{rgb}{0,0.5,0}

\DeclareRobustCommand{\Sec}[1]{Sec.~\ref{#1}}

\DeclareRobustCommand{\Fig}[1]{Fig.~\ref{#1}}

\DeclareRobustCommand{\Eq}[1]{Eq.~(\ref{#1})}

\DeclareRobustCommand{\Refs}[1]{Refs.~\cite{#1}}

\newcommand{\be}{\begin{equation}}
\newcommand{\ee}{\end{equation}}

\begin{document}

\title{An Indirect Model-Dependent Probe of the Higgs Self-Coupling}

\author{Matthew McCullough} 
\address{Center for Theoretical Physics, \\
Massachusetts Institute of Technology, \\
Cambridge, MA 02139, USA}

\date{\today}

\begin{abstract}
The Higgs associated production cross section at an $e^+ e^-$ collider is indirectly sensitive to the Higgs self-coupling, $h^3$, at next-to-leading order (NLO).  Utilizing this, a new indirect method is proposed for constraining deviations in the self-coupling below the di-Higgs production threshold in certain models.  Although this indirect constraint is model-dependent, making it valid only under specific assumptions, meaningful indirect constraints on the self-coupling may be realized.  Specific realistic scenarios where the indirect constraint applies are discussed and in particular it is shown that in the well-motivated class of two Higgs-doublet models there exist regions of parameter space in which the NLO effects from a modified self-coupling dominate over the LO modifications, demonstrating a concrete scenario in which large modifications of the Higgs self-coupling may be indirectly constrained using the proposed method.  Other models, such as strongly coupled scenarios, are also discussed.  The indirect method would give valuable constraints on deviations in the Higgs self-coupling, and would be complementary to the direct measurements possible with di-Higgs production at other colliders, providing precious additional information in the effort to unravel the properties of the Higgs boson.
\end{abstract}


\maketitle

\section{Introduction}
\label{sec:introduction}
The lack of evidence for beyond the Standard Model (BSM) physics at the LHC and the discovery of a SM-like Higgs \cite{Chatrchyan:2012ufa,Aad:2012tfa}, sharpens questions surrounding the hierarchy problem and new physics at the weak scale, and the recently discovered Higgs presents a unique opportunity to search for BSM physics.

Due to its gauge charges and spin the Higgs may interact with BSM fields and in many perturbative scenarios these interactions may modify couplings between the Higgs and SM fields at leading order (LO) either at tree level or loop level, and also at next-to-leading order (NLO) at loop level.  While the former case has received considerable attention, signals of NLO BSM Higgs  physics have yet to be fully explored, although preliminary investigations have shown great potential for unravelling the nature of the Higgs.  Evidence for BSM Higgs couplings may in fact arise at NLO, in some cases at greater significance than LO signals with $e^+ e^-$ colliders \cite{Englert:2013tya}.  NLO effects also allow for the resolution of the hierarchy problem, and naturalness of the weak scale, to be tested independent of the specifics of a particular model by constraining precisely the couplings to new fields which cancel quadratic divergences in the Higgs  mass \cite{Craig:2013xia}.  At the LHC significant modifications of the dependence of LHC Higgs observables on the SM Higgs couplings can arise once NLO effects are included \cite{Englert:2013vua}.  In this work a new application of BSM NLO Higgs physics is presented which enables a model-dependent indirect constraint on the Higgs self-coupling at energies below the di-Higgs production threshold.

Deviations of the self-coupling from the SM value may be parameterized with $\delta_h$, relating the true coupling to the SM value via $A_h = (1+\delta_h) A_{h,SM}$.  Techniques for directly measuring $\delta_h$ at both the LHC \cite{Baur:2002rb,Baur:2002qd,Baur:2003gp,Dolan:2012rv,Baglio:2012np} and future colliders \cite{Baer:2013cma,Asner:2013psa} have been pursued vigorously. The di-Higgs production rate is sensitive to the Higgs self-coupling through processes with an s-channel virtual Higgs, which also typically interfere with other di-Higgs production amplitudes such as di-Higgs box diagrams in gluon fusion at the LHC.  By observing the most promising di-Higgs final states it is possible to measure the Higgs self-coupling with precision estimates of $\delta_h \approx 50\%$ at a high luminosity LHC run, and from $21\%$ at the baseline of the ILC with staged running up to 1TeV, to $13\%$ with a luminosity upgrade \cite{Dolan:2012rv,Baglio:2012np,Goertz:2013kp,Barr:2013tda,Dawson:2013bba,Baer:2013cma,Asner:2013psa}.  It should also be kept in mind that these estimates are preliminary, conservative, and luminosity-limited.  Thus considerable improvement in these estimates may be achieved with further study and/or modified collider running strategies.

It is worth emphasizing that these scenarios measure the Higgs self-coupling directly at LO and hence require a CM energy $\sqrt{s} > 2 m_h$, and in the case of an $e^+ e^-$ collider $\sqrt{s} > 2 m_h + m_Z$.  For this reason it is typically assumed that it is impossible to gather information on the Higgs self-coupling with an $e^+ e^-$ collider operating below the di-Higgs threshold, such as the proposed TLEP collider which would operate at high luminosity at lower energies $\sqrt{s} < 2 m_h + m_Z$ and only at reduced luminosity above the di-Higgs production threshold \cite{Gomez-Ceballos:2013zzn}.

In this work, by exploiting NLO rather than LO effects, it is demonstrated through a one loop calculation in \Sec{sec:introduction} that in the context of certain models it is possible to constrain modifications of the Higgs self-coupling indirectly at an $e^+ e^-$ collider through precision measurements of the Higgs associated production cross section.  For example, as will be discussed in \Sec{sec:onecoup}, for the proposed TLEP parameters \cite{Gomez-Ceballos:2013zzn} running at $\sqrt{s} \sim 240$ GeV it would be possible to constrain deviations in the Higgs self-coupling indirectly at  an accuracy of $28\%$, under the model-dependent assumption that only the Higgs self-coupling is modified.

By extending the assumed parameter modifications by only one parameter to include a modification to the $hZZ$ vertex by a constant energy-independent factor, and momentarily assuming that that possible energy-dependent modifications vanish, then this LO modification alone would typically swamp the NLO effect from a modified Higgs self-coupling.  In this case a measurement of the associated production cross section at $\sqrt{s}=240$ GeV, $\delta_\sigma^{240}$, can constrain a linear combination of the deviations in the self-coupling, $\delta_h$, and also the deviation in the $hZZ$ coupling, $\delta_Z$, as
\be
\delta_\sigma^{240} = 100 \left(2 \delta_Z + 0.014 \delta_h \right) \% ~~,
\label{eq:effectiveth}
\ee
but not the self-coupling alone.  Thus in order to set a constraint on $\delta_h$ from a single measurement it would be necessary to make additional assumptions on $\delta_Z$.  In this particular case in \Sec{sec:twocoup} it is shown that combinations of precision associated production measurements at different center of mass energies may be used to determine ellipse-plot constraints on the combined parameter space of $\delta_Z$ and $\delta_h$, which could be used to set constraints on some strongly-coupled Higgs models.

Continuing to study specific model scenarios it is interesting to consider whether there are any renormalizable UV-complete models where it can be demonstrated that the NLO effects of a modified self-coupling may dominate over the possible LO effects from a modified $hZZ$ coupling.  In \Sec{sec:2HDM} it is shown that such a scenario in fact arises in the decoupling limit of a two Higgs-doublet model (2HDM).  In these models in the decoupling limit the modification of the $hZZ$ coupling scales approximately as $\delta_Z \sim v^4/m_A^4$, where $v$ is the electroweak breaking Higgs vacuum expectation value and $m_A$ is the mass of the additional pseudoscalar in a 2HDM.  On the other hand the self-coupling modification decouples less rapidly as $\delta_h \sim v^2/m_A^2$.  Due to this, for $m_A \gtrsim 750$ GeV the additional NLO loop factor in the self-coupling modification of the associated production cross section is larger than the additional factor of $v^2/m_A^2$ suppressing the LO modification of the $hZZ$ vertex, and the self-coupling NLO modification in fact dominates over the LO modification.  Thus in this parameter range in the well-motivated class of 2HDMs the NLO effect described here may be used to set indirect constraints on the Higgs self-coupling.  

Finally, in \Sec{sec:modind} more general, model-independent scenarios are discussed.  Typically a large number of different energy dependent deviations may enter the associated production cross section\footnote{I am grateful to two anonymous referees for bringing this to my attention.} and contrive to cancel effects between each other in the final cross section, meaning that in a truly model-independent sense it is not possible to extract an unambiguous constraint on the self-coupling in this way.  This is a general weakness of indirect constraints on higher dimension operators and the usual caveats about various different contributions from different operators canceling in the final result are discussed.  This also demonstrates that an indirect constraint cannot unambiguously single out a modified Higgs self-coupling as the cause of a deviation in the cross section measurement.  Nonetheless, subject to these caveats, this indirect constraint could be used to place interesting bounds on deviations of the Higgs self-coupling, and would give invaluable information complementary to the direct measurements possible at other colliders.  Conclusions are presented in \Sec{sec:conc}.

\section{The One-Loop Correction}
\label{sec:introduction}
In studies aimed at measuring the Higgs self-coupling through di-Higgs production it is often assumed that all other Higgs couplings take SM values and the Higgs is not coupled to any new BSM fields.  This is a useful assumption since a number of different Higgs couplings, and fields, enter the di-Higgs production process, leading to some degeneracy between the effects of a modified Higgs self-coupling and other modified Higgs couplings.  Solely for calculational simplicity this simplifying assumption is employed in this section and readers are directed to \Sec{sec:theory} for a discussion of the relevant assumptions in theoretically realistic scenarios.  The interactions are given by the following Lagrangian
\be
\mathcal{L} = \mathcal{L}_{SM} - \frac{1}{3!} \delta_h A_{h,SM} h^3 ~~.
\ee
Such a modification can arise from the following nonrenormalizable addition to the Higgs potential
\be
V_h = V_{h,SM} + \frac{1}{\Lambda^2} \left(v^2-|H|^2 \right)^3 ~~,
\label{eq:pot} 
\ee
where the scale $\Lambda$ is associated with the scale of new physics in the Higgs sector, such as the mass scale of new fields or the scale of strong dynamics.  This modification enters the calculation of Higgs processes at LO and NLO.  \Eq{eq:pot} shows that scenarios which are purely SM-like with the exception of non SM-like Higgs self-couplings are in fact completely consistent with electroweak symmetry in the UV.  Thus no pathologies related to the underlying gauge symmetry will arise with a modified self-coupling.  If processes involving the Higgs self-coupling at tree-level are considered, such as in di-Higgs production, then the modified coupling can be simply included in LO calculations.  However if an NLO calculation encounters the Higgs self-coupling at LO and at NLO, as in di-Higgs production, then a suitable counter-term for the irrelevant operator in \Eq{eq:pot} must be calculated following procedures for loop calculations in effective field theories \cite{Weinberg:1978kz}.  In processes where the Higgs self-coupling does not contribute at LO but does enter at NLO, as in the single Higgs production considered here, the modified self-coupling can be included in one-loop diagrams without recourse to the details of renormalization of the irrelevant operator in \Eq{eq:pot}, however proceeding to NNLO in this case would require the counter-term to this operator.

\begin{figure}[]
  \centering
  \includegraphics[height=0.175\textwidth]{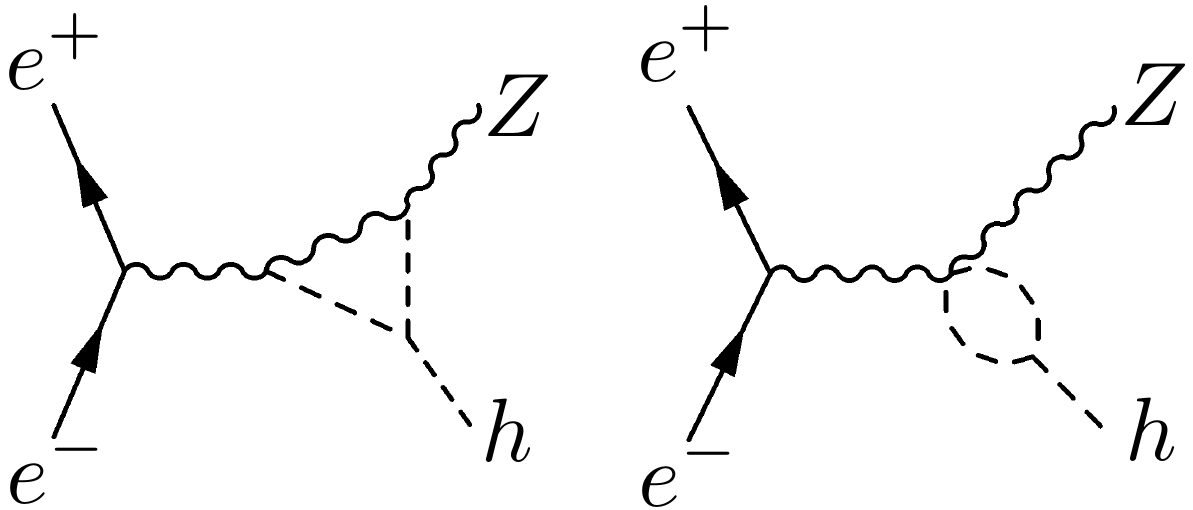}
  \caption{NLO vertex corrections to the associated production cross section which depend on the Higgs self-coupling.  These terms lead to a linear dependence on modifications of the self-coupling $\delta_h$.}
  \label{fig:vertex}
\end{figure}

The dominant Higgs production process at an $e^+ e^-$ collider at the energies considered here is Higgs associated production.  At NLO the Higgs self-coupling enters the associated production amplitude in two ways.  It enters quadratically via a modified Higgs wavefunction counter-term, feeding into associated production at NLO as a modification of the $hZZ$ coupling.  The self-coupling also enters into the amplitude linearly through diagrams such as \Fig{fig:vertex}.  Depending on gauge choice there are also diagrams with internal Goldstone lines.

The full NLO corrections to $e^+ e^- \to hZ$ are determined using the {\sc{FeynArts}}, {\sc{FormCalc}}, and {\sc{LoopTools}} suite of packages \cite{Hahn:2000kx,Hahn:1998yk} by calculating the full one-loop electroweak corrections to associated production (see \Refs{Fleischer:1982af,Jegerlehner:1983bf,Fleischer:1987zv,Denner:1992bc}) and extracting the dependence on the self-coupling parameter.  The counter-terms for all SM-Higgs couplings are calculated automatically following the electroweak renormalization prescription of \cite{Denner:1991kt}.  The analytic form of the correction at a CM energy $\sqrt{S}$ can be extracted from the {\sc{FeynArts}} and {\sc{FormCalc}} \cite{Hahn:2000kx,Hahn:1998yk} output in terms of the various one-loop integrals
\be
B (p^2,M_1^2,M_2^2)   =  \int \frac{K d^D q}{[q^2-M_1^2] [(q+p)^2-M_2^2]} ~,
\ee
and
\begin{eqnarray}
 C_{\mu_1,..,\mu_N} (k_1^2,(k_1-k_2)^2,k_2^2,M_1^2,M_2^2,M_3^2)  =   \nonumber \\ 
 \int \frac{K q_{\mu_1} \cdot \cdot \cdot q_{\mu_N} d^D q}{[q^2-M_1^2] [(q+k_1)^2-M_2^2] [(q+k_2)^2-M_3^2]} ~,
\end{eqnarray}
where
\be
K=\frac{\mu^{4-D}}{i \pi^{D/2} r_\Gamma} ~, ~~~~ r_\Gamma = \frac{\Gamma^2 (1-\epsilon) \Gamma (1+\epsilon)}{\Gamma (1-2 \epsilon)} ~.
\ee
The two-point scalar function encountered here is defined as
\be
B_0 = B (M_H^2,M_H^2,M_H^2),
\ee
and the first derivative of this function as
\be
B_0' = \partial B (p^2,M_H^2,M_H^2)/\partial p^2 |_{p^2=M_H^2} ~.
\ee
The three-point scalar functions are
\be
C_0=C (M_H^2,S,M_Z^2,M_H^2,M_H^2,M_Z^2),
\ee
and $C_1$, which is the scalar coefficient of $k_1$ in $C_{\mu_1}$ with the same arguments.  $C_{00},C_{11},C_{12}$ are the scalar coefficients of $g_{\mu,\nu},k_1 k_1,$ and $k_1 k_2$ in $C_{\mu_1,\mu_2}$.  All of these functions can be easily evaluated using the {\sc{LoopTools}} package \cite{Hahn:2000kx,Hahn:1998yk}.  With these definitions the full form of the self-coupling correction is
\begin{eqnarray}\label{eq:allterms}
\delta_\sigma (S) &  = & \frac{\sigma_{\delta_{h}\neq0}}{\sigma_{\delta_{h}=0}} -1  \\
& = &  \frac{3 \alpha M_H^2 \delta_h}{16 \pi \sin(\theta_W)^2 M_W^2 \beta} \times \nonumber \\
 && \text{Re} \bigg[2 \left(S+M_Z^2- M_H^2\right) (12 M_Z^2 S - \beta) \kappa  - \zeta \beta \bigg] , \nonumber
\end{eqnarray}
where
\begin{eqnarray}
\beta & = & (M_H^2-M_Z^2)^2+10 M_Z^2 S+S^2-2 M_H^2 S, \\
\zeta & = & B_0-4 C_{00}+4 C_0 M_Z^2+3B_0' M_H^2
\end{eqnarray}
and
\be
\kappa =C_1+C_{11}+C_{12}.
\ee
\Eq{eq:allterms} was calculated in the $R_\xi$ gauges, and the absence of the $\xi$ parameter demonstrates the full gauge invariance of the result.  Furthermore, although a number of UV-divergences appear individually, the final result is UV-finite as these divergences cancel in $B_0-4 C_{00}$ and also in $\kappa$.

\begin{figure}[]
  \centering
  \includegraphics[height=0.32\textwidth]{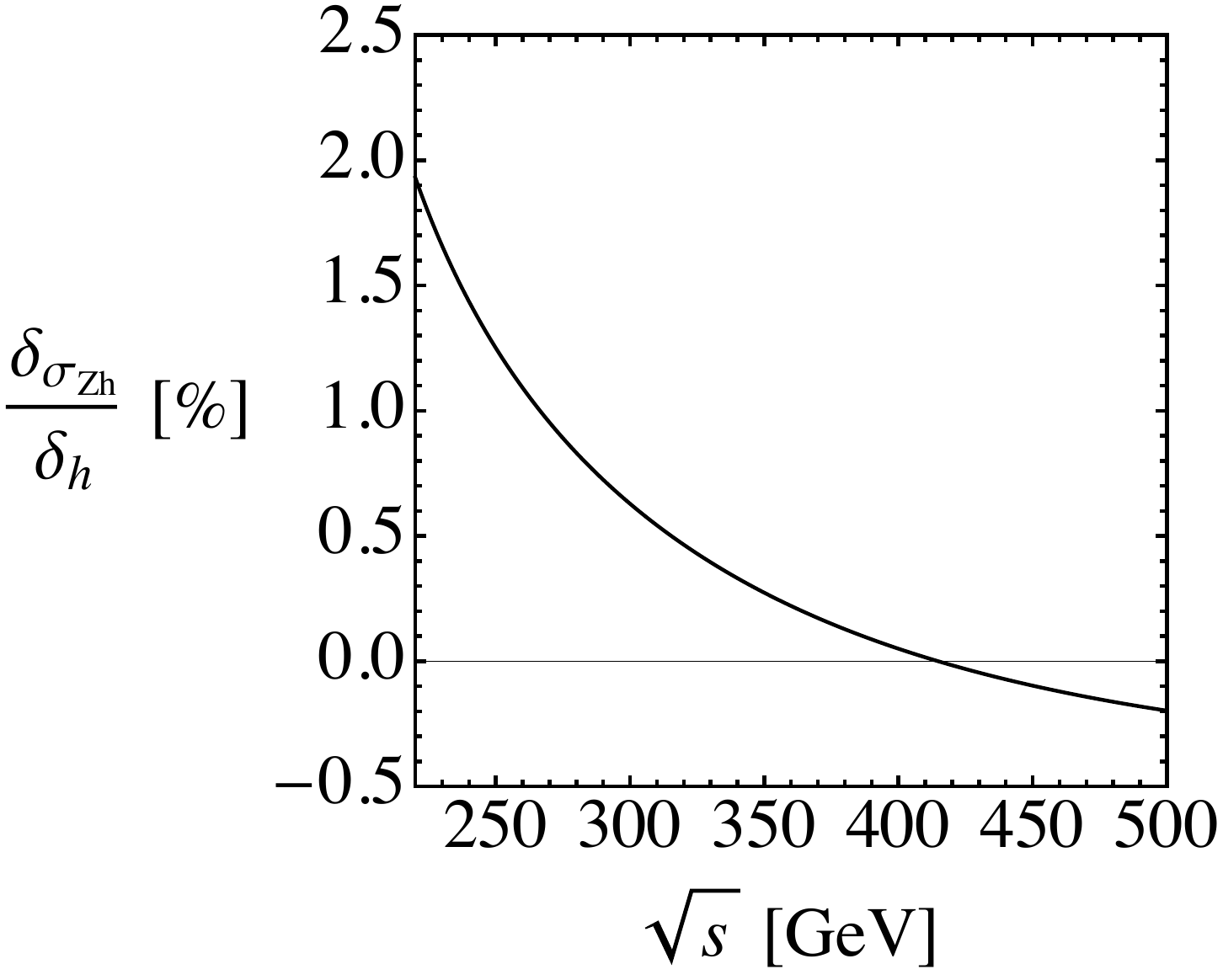}
  \caption{Corrections to $\sigma (e^+ e^- \to hZ)$, for a given variation in the self-coupling, $\delta_h$, as a function of the CM energy from $220$ to $500$ GeV.}
  \label{fig:edep}
\end{figure} 

At various CM energies the fractional corrections to the associated production cross section, $\sigma_{\delta_{h}} (e^+ e^- \to hZ)$, relative to the SM rate are found to be
\be
\delta_\sigma^{240,350,500}  = 1.4,0.3,-0.2 \times \delta_h  \% ~~,
\label{eq:result}
\ee
where only the lowest-order term in $\delta_h$ has been retained as other higher-dimension operators may contribute at $\mathcal{O}(\delta_h^2)$, and the coefficient of this term is unknown.  The full energy dependence is shown in \Fig{fig:edep}. 

\section{Model-Specific Applications}
\label{sec:theory}
Although it is interesting to consider the role of the Higgs self-coupling in NLO electroweak corrections and compare the magnitude of these corrections to the projected experimental sensitivity at future Higgs factories, it is important to consider the interpretation of this result with regard to realistic theoretical scenarios in which the self-coupling is modified.

When discussing the experimental sensitivity of certain measurements to specific higher dimension operators or modified couplings there is some precedent, in many fields including Higgs physics, for considering only the effects of modifying a single operator or coupling at a time when discussing experimental sensitivity.  Phrased in this way, only considering a modified self-coupling, then at a $240$ GeV run of TLEP the associated production cross section precision would be $0.4 \%$ \cite{Gomez-Ceballos:2013zzn} with $10 {\text{ab}}^{-1}$, and this would give sensitivity to deviations of the Higgs self-coupling, $\delta_h$, at the level of $|\delta_h| \lesssim {28 \%}$.  However, this sensitivity is highly dependent on the specific assumption that only the self-coupling has been modified.  In particular in most cases the effects from a number of different higher dimension operators or modified couplings may in fact conspire and interfere against one another to produce a measurement consistent with the SM prediction, even in the presence of significant underlying modifications.\footnote{For example in discussions of modified Higgs couplings the SM coupling is usually rescaled with one parameter and fitted to the data.  However in reality such couplings may be modified in a number of ways, by multiple higher dimension operators with different energy dependence, and only linear combinations of the modifications, which may interfere with one another, may be constrained in each case.}  Thus, as with any indirect probe of new physics, any statement of the experimental sensitivity comes with implicit assumptions and it is important to consider the model-dependence and specifically the possibility of multiple coupling modifications and interference between their effects.  This is particularly necessary in this case where the NLO effect of the self-coupling may be overwhelmed by LO modifications.

In this section various new physics scenarios are considered and the interpretation of the indirect constraint is discussed.

\subsection{Re-Scaled $h^3$ SM-Coupling}\label{sec:onecoup}
The original assumption that all Higgs couplings are SM-like with the exception of the Higgs self-coupling is now considered.  It may be the case that this specific scenario arises, however it would be necessary that no higher-dimension operators involving gauge fields are generated and also that the new physics only modifies the Higgs potential and not kinetic terms.  Such a scenario is thus unlikely from an effective theory perspective, however it may arise in certain models, or it may be the case that the modification of the self-coupling is far greater than other coupling modifications.  An explicit example realizing the latter possibility is described in \Sec{sec:2HDM}.

In the scenario with only a modified self-coupling, under one-loop RG evolution the dimension six operator \Eq{eq:pot} does not generate any additional operators which lead to additional modifications of the other Higgs couplings \cite{Jenkins:2013zja,Jenkins:2013wua,Alonso:2013hga}.  It should be noted that this statement only holds for dimension six operators and only at one loop.  RG evolution will generate modifications in the other Higgs couplings which enter into associated production at two-loops, however these additional RG-contributions would subdominant even though they modify the tree-level $hZZ$ coupling.  Thus if, and only if, the model in question is defined as
\be
V_h = V_{h,SM} + \frac{1}{\Lambda^2} \left(v^2-|H|^2 \right)^3 ~~,
\ee
then it would be possible to determine constraints on deviations of the Higgs self-coupling, $\delta_h$, in this specific model at the level of $|\delta_h| \lesssim {28 \%}$, and this constraint is robust against RG evolution.

\subsection{Re-Scaled $h^3$ and $hZZ$ SM-Couplings}
\label{sec:twocoup}
In generic new physics scenarios additional operators which modify the tree-level $hZZ$ coupling will often be generated, and they would generically dominate over the NLO modification due to the self-coupling, significantly complicating the interpretation of any indirect constraint.   There are a number of operators which may modify the $hZZ$ coupling, however in this section it is assumed that no new operators involving the gauge fields are generated and that custodial symmetry is respected.  In this case the operator of \Eq{eq:pot} and
\be
\mathcal{L}_{Eff} = \frac{1}{\Lambda_H^2} \partial^\mu |H|^2 \partial_\mu |H|^2 + ...,
\label{eq:kin}
\ee
give the leading corrections in strongly coupled Higgs scenarios \cite{Giudice:2007fh,Contino:2013gna}.\footnote{Similarly, at dimension $8$ there are only two operators generated \cite{Contino:2013gna}, which lead to similar modifications of the $hZZ$ and $h^3$ couplings.}  In the electroweak breaking vacuum and after performing a field rescaling for a canonically normalized Higgs kinetic term, \Eq{eq:kin} modifies all Higgs couplings, and in particular the $hZZ$ coupling by a constant factor $\delta_Z \sim v^2/ \Lambda_h^2$.  Using naive dimensional analysis (NDA) \cite{Manohar:1983md,Georgi:1985kw,Georgi:1986kr,Georgi:1992dw} for a strongly coupled Higgs scenario the modifications of the self-coupling would be a factor $\sim (4 \pi)^2$ larger than the modifications of the $hZZ$ coupling.  Thus in this case it would be expected that the deviation in the associated production cross section from a modified $hZZ$ coupling at tree level would be of a similar magnitude to the loop-level effect from modified self-coupling.\footnote{See e.g.\ \cite{Grinstein:2007iv} for an explicit example where this would be the case.}  However for clarity in this work the loop-suppression of the deviation from the self-coupling will be explicitly written and the NDA factors will not be included.

This type of scenario where the SM Higgs couplings, in this case $hZZ$ and $h^3$, are rescaled by some common factor is often considered in modified Higgs coupling analyses rather than considering the effects of higher dimension operators, making this section analogous to these re-scaled coupling scenarios.  Now including these modifications, and taking the leading-order coefficients of $\delta_Z$ and $\delta_h$ and only expanding to first order in any $\delta$, the associated production cross-section would vary as
\be
\delta_\sigma^{240} = 100 \left(2 \delta_Z + 0.014 \delta_h \right) \% ~~,
\label{eq:effectiveth}
\ee
Thus in this specific model a single precision measurement of the associated production cross section can constrain this linear combination of couplings.  Also, if $\delta_Z \sim \delta_h$, as would typically be expected in perturbative scenarios, the LO modification of the associated production cross section from $\delta_Z$ would completely dominate the NLO modification from $\delta_h$.

\begin{figure}[]
  \centering
  \includegraphics[height=0.42\textwidth]{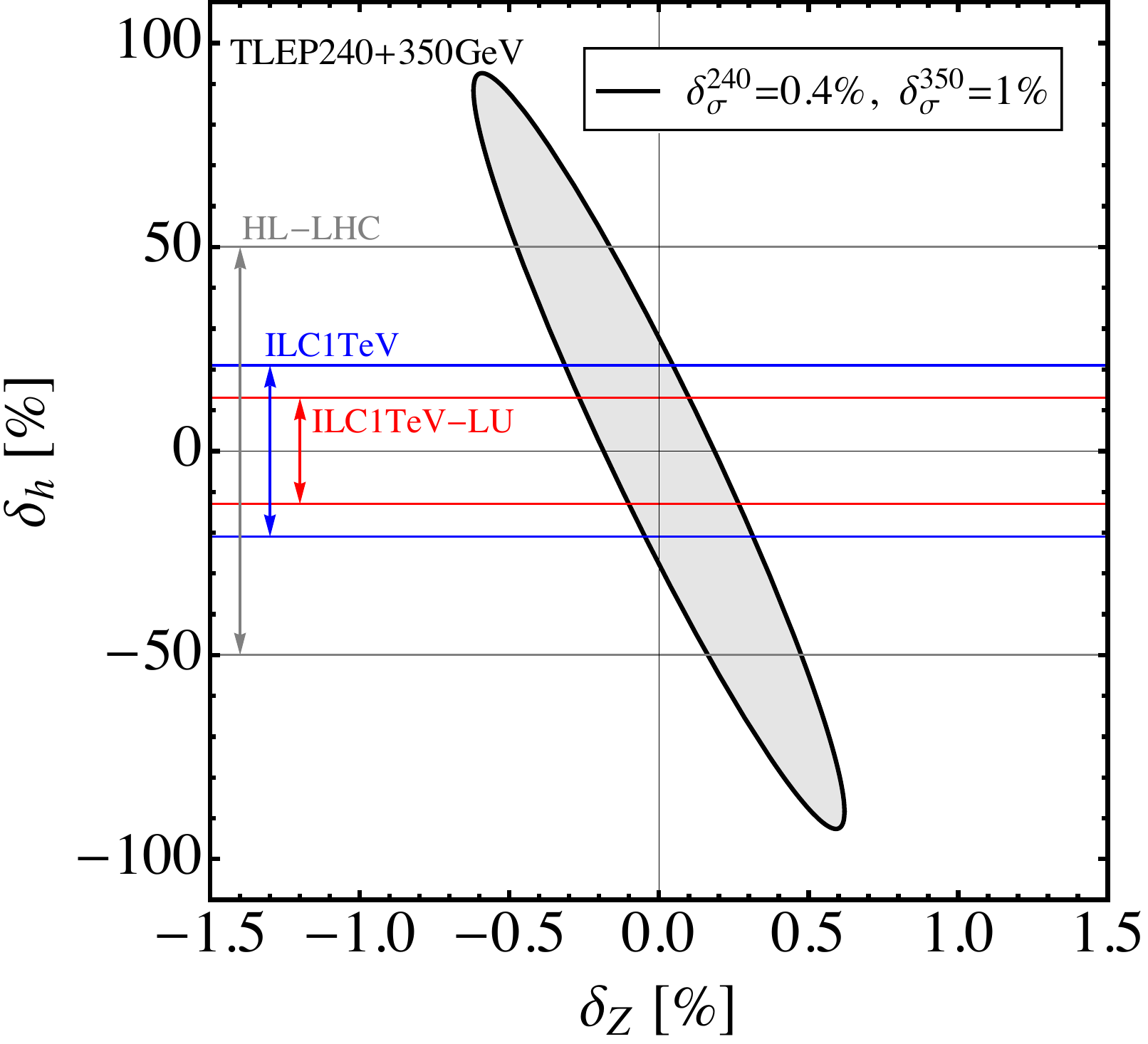}
  \caption{Indirect $1\sigma$ constraints possible in $\delta_Z - \delta_h$ parameter space by combining associated production cross section measurements of $0.4\%$ ($1\%$-estimated) precision at $\sqrt{s} = 240$ GeV, ($350$ GeV) in solid black.  For large values of $|\delta_h|$ this ellipse can only be considered qualitatively as the calculation is only valid to lowest order in $\delta_h$.  The different scales should be noted.  Direct constraints possible at the high luminosity LHC and 1 TeV ILC (with LU denoting luminosity upgrade) are also shown for comparison.  This plot only applies to the specific model discussed in \Sec{sec:twocoup} and if energy-dependent $hZZ$ couplings were allowed then such a constraint could not be determined.}
  \label{fig:ellipse}
\end{figure}

However, from \Eq{eq:result} it is clear that the NLO self-coupling correction is energy-dependent, meaning that measurements at different energies constrain different linear combinations of coupling modifications, which may lead to ellipse-plot constraints in the space of $\delta_Z - \delta_h$ couplings.\footnote{Similar multiple-energy measurements have been proposed to disentangle the effects of $hhZZ$ and $h^3$ modifications in di-Higgs production at the ILC \cite{Contino:2013gna}.}  In \Fig{fig:ellipse} the indirect ellipse constraint that would result from precision measurements at $240$ GeV and $350$ GeV is shown.  A cross section precision of $0.4\%$ at $240$ GeV has been assumed \cite{Gomez-Ceballos:2013zzn}.  Studies of the cross section precision at $350$ GeV have not yet been performed, and a rough estimate of $1\%$ precision has been assumed here.  This ellipse only applies to the specific model assumptions employed in this section, but demonstrates that under the assumption of a rescaled $hZZ$ coupling and Higgs self-coupling interesting constraints may be imposed on deviations of both parameters, with relevance to strongly coupled Higgs scenarios.

\subsection{Two Higgs-Doublet Scenarios}\label{sec:2HDM}
Precision measurements of Higgs associated production at a lepton collider may play an important role in constraining the Higgs self-coupling in two Higgs-doublet models (2HDMs).  In 2HDMs there are a number of free parameters which determine the couplings of the SM-like Higgs boson to other fields.  This section will only be concerned with the couplings to SM fields, which, in a CP-conserving 2HDM, may be parameterized with $\alpha$, $\beta$, and the pseudoscalar mass $m_{A}$.\footnote{For simplicity it is assumed that the 2HDM couplings such as $|H_1|^2 H_1 \cdot H_2^\dagger$ are set to zero.  Including these couplings does not change the conclusions of this section.}  Assuming that the observed SM-like Higgs boson is the lightest CP-even scalar of the 2HDM and making the replacement $\cos (\beta-\alpha) = \delta$, which measures the deviations of the Higgs couplings from the SM values, then in terms of these parameters the tree-level Higgs coupling to the Z-boson is modified from the SM value to
\be
1+\delta_Z = \sin(\beta-\alpha) =  \sqrt{1-\delta^2} ~~,
\ee
and the Higgs self-coupling is modified from the SM value by the factor
\begin{eqnarray}
1+\delta_h & = &  \sqrt{1-\delta ^2} \left(1+2 \delta ^2\right)+2 \delta ^3 \cot (2 \beta ) - \nonumber \\ 
&& 2 \delta ^2 \frac{m_{A}^2}{m_h^2} \left(\delta  \cot(2 \beta )+\sqrt{1-\delta ^2} \right) ~~.
\label{eq:2HDMdh}
\end{eqnarray}

\begin{figure}[]
  \centering
  \includegraphics[height=0.43\textwidth]{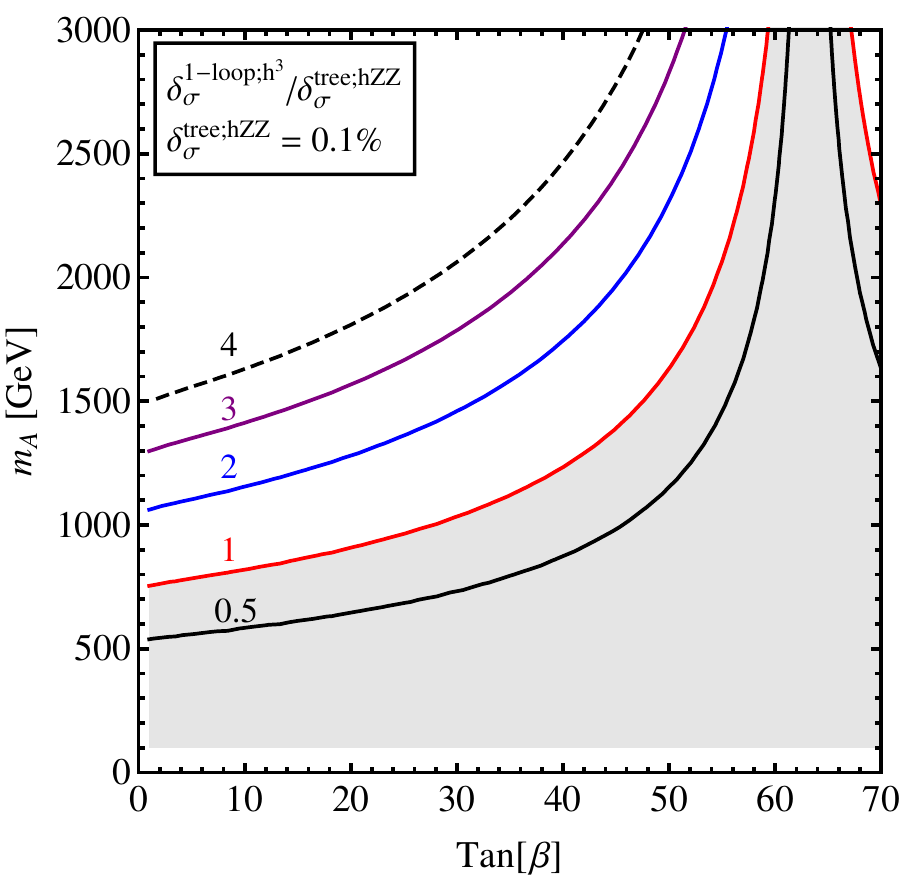}
  \caption{Contours of the ratio of NLO modifications to $\sigma (e^+ e^- \to h Z)$ from a modified Higgs self-coupling relative to the LO modifications due to the modified $hZZ$ vertex in a 2HDM as a function of the parameters $\beta$ and $m_{A}$.  In the gray region the LO modifications dominate and in the white region the NLO corrections involving the self-coupling dominate.  Loops of additional heavy scalars are not included and are estimated to be subdominant.  The LO $hZZ$ coupling modification is set to a constant value of $0.1\%$, such that the region above the dashed line corresponds to deviations greater than the expected experimental sensitivity.  For fixed $\delta$ and large $m_{A}$ we have $\delta \sim \overline{\lambda} v^2 / 2 m_A^2$ \cite{Englert:2014uua}, where $\overline{\lambda}$ is a combination of dimensionless couplings and mixing angles in the Higgs sector, thus large $m_A$ and $\delta^2 = 0.1 \%$ large requires almost non-perturbative couplings.  This does not, however, alter the ratio of the magnitude of effects from the NLO self-coupling modification relative to the LO $hZZ$ modifications, and large or non-perturbative couplings are not required for the self-coupling modification to dominate.}
  \label{fig:e2HDM}
\end{figure}
The second line of \Eq{eq:2HDMdh} demonstrates that in a generic 2HDM the modifications to the Higgs self-coupling may be large, and in cases where $m_A \gg m_h$ they are typically larger than the modifications to the $hZZ$ coupling due to the $m_A^2/m_h^2$ enhancement of the $\delta^2$ term. The deviations in the self-coupling may not be arbitrarily large as the self-coupling still obeys the decoupling property in the large $m_A$ limit $\delta \propto v^2/m_A^2$ \cite{Gunion:2002zf}, thus the total deviation still decouples as $\delta_h \propto v^2/m_A^2$.   However, the quantity of interest here is the ratio of deviations due to the self-coupling relative to the deviations from the modified $hZZ$ coupling, and it has recently been emphasized that $\delta_h/\delta_Z \approx 4 m_A^2/m_h^2$ \cite{Efrati:2014uta}, thus it is typical in a 2HDM for the modification of the self-coupling to be greater than the modification of the $hZZ$ vertex, particularly in the decoupling limit $m_A \gg m_h$.  This raises the possibility of the loop-level modifications to the associated production cross section involving the Higgs self-coupling exceeding the tree-level modifications from the $hZZ$ vertex in this class of models.   From \Eq{eq:effectiveth} we see that this occurs if $0.014 \delta_h > 2 \delta_Z$, and from the previous relation it is clear that for approximately $m_A \gtrsim 750$ GeV this is the case.

In \Fig{fig:e2HDM} contours are shown of the ratio of associated production cross section modifications from the Higgs self-coupling at one-loop divided by the tree-level modifications due to the modified $hZZ$ vertex at tree-level in a 2HDM.  In the gray shaded region the tree-level modifications dominate, and in the white region the loop-level self-coupling modifications dominate.

For comparison with experimental prospects the modification of the $hZZ$ vertex is set to $\delta^2 = 0.1\%$, thus above the dashed line the deviations in the associated production cross section due to the modified Higgs self-coupling become comparable to the expected experimental sensitivity.  The funnel-like feature in \Fig{fig:e2HDM} can be understood as in the limit $m_A \gg m_h$ the self-coupling correction is dominated by the second term of \Eq{eq:2HDMdh} and this second term vanishes for $\tan (\beta) \approx 2/\delta$, independent of the pseudoscalar mass.

To fully understand all relative contributions to the associated production cross section a complete calculation, which is beyond the scope of this work, would also include loops of heavy scalars.  However in regions with large $m_A$ the corrections from loops of heavy scalars would likely be subdominant as although factors proportional to $m_{A}^2/m_h^2$ may appear in scalar vertices, the loop integrals would also decouple with increasing $m_A$, unlike the Higgs self-coupling loops of \Fig{fig:vertex}, meaning that the modification from the Higgs loops with modified self-coupling would dominate over the heavy scalar loops.

This is an explicit demonstration of the existence of a well-motivated perturbative model where modifications of the self-coupling may lead to deviations in the associated production cross section from NLO effects which are observable and dominate over the LO deviations from the modified $hZZ$ vertex.  The indirect constraint on the self-coupling proposed here would be very useful for constraining the self-coupling in this realistic and commonly studied example.  Furthermore, this also demonstrates that if only LO coupling modifications are assummed the precision constraints on scenarios such as 2HDMs could be misinterpreted, and the opportunity to learn much more about the structure of such models through NLO effects, including effects due to the self-coupling, would be missed.

\subsection{Generic New Physics Scenarios}\label{sec:modind}
Model-independent scenarios are now finally considered.  It is possible to capture the effects of generic new physics scenarios by allowing all higher dimension operators consistent with the gauge symmetries of the SM.  A number of operators which modify the tree-level $hZZ$ coupling arise at dimension six and have varying energy dependence, increasing the list of undetermined parameters.  Due to these operators any precision associated production cross section measurement would constrain a linear combination of all of these parameters, including the self-coupling, and an unambiguous extraction of limits on the self-coupling is not possible.  This is true for many measurements of Higgs properties and also for many indirect constraints on higher dimension operators in high energy physics.  This case can be shown schematically as
\begin{eqnarray}
\delta_\sigma (S)  & \sim & 2 \sum_a \kappa_a (S) \delta_{hZZ,a} + (\text{loop-factor}) \delta_h \nonumber \\
&& + (\text{loop-factor}) \sum_b  \kappa_b (S) \delta_{hXX,b}   ~~,
\label{eq:scheme}
\end{eqnarray}
where the $\kappa(S)$ are various energy-dependent coefficients coming from LO modifications of the $hZZ$ vertex and the $\delta_{hXX,b}$ are effectively coefficients of various higher dimension operators which modify the Higgs couplings to all fields and enter at NLO.  Furthermore, it would typically be expected that the modifications due to the tree-level operators would dominate, thus if $|\delta_{hZZ,a}| \lesssim \mathcal{O} (1 \%)$ it would seem unlikely that $|\delta_{h}| \sim \mathcal{O} (28 \%)$.  However, in a model-independent study all of the coefficients should be allowed to vary independently, as there may be an underlying scenario in which $|\delta_{h}| \gg |\delta_{hZZ,a}|$, such as in \Sec{sec:2HDM}.  By allowing this variation then no truly model-independent constraint is possible, as with many indirect constraints on higher dimension operators.  Assumptions regarding the possible cancellations between different terms may be imposed, which essentially amounts to imposing some degree of model dependence.

The degree of cancellation assumed between different operators may be quantitatively understood through
\be
\Delta = \frac{\delta_\sigma (S)}{\delta_{\sigma_h} (S)} ~,
\ee
which is the measured deviation in the total cross-section relative to the total contribution from the self-coupling.  For a measurement of the associated production cross section which is purely SM-like (i.e.\ $\delta_\sigma^{240} =0$) with an accuracy of $0.4\%$, then if it is assumed that cancellations between contributions may only occur at the level of $\Delta \%$ the model-dependent constraint on the self-coupling would be
\be
|\delta_{h}| < 28 \% \times \frac{100\%}{\Delta (\%)} ~.
\ee
If cancellations between various contributions were tolerated at the $25\%$ level the constraint is $|\delta_{h}| < 114 \%$ and so on.  This information would still be very valuable for understanding the possible deviations in a number of couplings, including the self-coupling, but the necessity of an assumption about possible cancellations also demonstrates that the greatest utility of the indirect method for constraining the self-coupling is in the context of specific models, such as the 2HDM discussed in \Sec{sec:2HDM} or potentially in some strongly coupled scenario as discussed in \Sec{sec:twocoup}.

\section{Conclusions}\label{sec:conc}
A method for indirectly constraining deviations in the Higgs self-coupling has been proposed and explored.  If it is assumed that only the self-coupling has been modified, an $e^+ e^-$ synchrotron such as TLEP operating at $240$ GeV can indirectly constrain deviations in this coupling at the level of $|\delta_h| \lesssim 28\%$.  If, in addition, it is assumed that the SM $hZZ$ coupling has also been modified then measurements at multiple energies may be combined to determine an ellipse-plot constraint in the two-dimensional parameter space of coupling modifications.  It has also been demonstrated that in 2HDMs in the decoupling limit NLO deviations in the associated production cross section from a modified self-coupling can dominate over the LO deviations from a modified $hZZ$ coupling, and the indirect method proposed here could be used to constrain modifications of the self-coupling in this scenario.   This demonstrates an application of this method to a well-motivated and commonly studied perturbative scenario.

In the case of completely generic model-independent new physics scenarios a number of higher-dimension operators may enter the associated production process and may interfere with the contribution from the modified self-coupling.  As with any indirect constraint on new physics this weakens any interpreted indirect constraint on the self-coupling from a precision cross section measurement due to the possibility of cancellations between different operators.  However, as with other indirect constraints on new physics scenarios, it is still possible to extract information on the self-coupling if it is assumed that no cancellations between various contributions are occurring.  

The proposed indirect constraint is not equivalent to a direct measurement at the LHC or ILC, as the different types of measurements constrain different linear combinations of possible coupling deviations and are thus subject to different model-dependent assumptions, with the indirect constraint arguably suffering from the greatest model-dependence.  However, this indirect information would be extremely valuable and, as it would constrain combinations of coupling deviations orthogonal to the direct measurements, would provide rare complementary insight into the structure of the Higgs potential, which may be the last frontier of a future precision Higgs program.

\section{Note Added}
After the publication of this work the author became aware of additional relevant literature.  Analogously to the present study, in the LEP era there was a program focussed on indirectly constraining modified gauge boson self-couplings through their influence on LEP precision measurements at one-loop, even in the presence of operators which modify such processes at tree-level (see e.g.\ \cite{DeRujula:1991se,Burgess:1992va,Burgess:1992gx,Hagiwara:1992eh,Hagiwara:1993ck,Hernandez:1993pp,Aihara:1995iq,Gounaris:1995mx,Brunstein:1996fz,Alam:1997nk,GonzalezGarcia:1999fq,Molnar:1999fk}).  For example, in \cite{Hagiwara:1993ck} indirect constraints were placed on four operators which enter at tree-level and five operators which enter only at one-loop each in isolation by assuming that all other operators, including tree-level modifications, were set to zero (see Table 1 of \cite{Hagiwara:1993ck} and also Table 1 and the surrounding discussion in \cite{Alam:1997nk}). These references also include detailed discussions highlighting the pitfalls and model-dependence of such indirect constraints as well as the necessity for directly probing such operators in tree-level processes such as gauge boson scattering.  Thus there are strong analogies between these past studies and the current work and \Refs{DeRujula:1991se,Burgess:1992va,Burgess:1992gx,Hagiwara:1992eh,Hagiwara:1993ck,Hernandez:1993pp,Eboli:1994jh,Aihara:1995iq,Gounaris:1995mx,Brunstein:1996fz,Alam:1997nk,Eboli:1998hb,GonzalezGarcia:1999fq,Molnar:1999fk} provide useful context and perspective.

\section*{Acknowledgements}
I am grateful for conversations with Nathaniel Craig, Sally Dawson, Christoph Englert, Patrick Fox, Markus Klute, Yann Mambrini, Tilman Plehn, Matthew Reece, Jesse Thaler, Junping Tian, and Michael Trott, for useful comments from anonymous referees, and to Tao Liu and Chen Shen for pointing out a typo in \Eq{eq:allterms} and cross checking results.
\bibliography{HiggsSelfLCref}

\end{document}